\documentclass{aa}
\include{epsf}

\begin{document}

\title{
Flux-loss of buoyant ropes interacting with convective flows
}

\titlerunning{Flux-loss of buoyant ropes}

\author{S.B.F. Dorch\inst{1}
\and B.V. Gudiksen\inst{1}
\and W.P. Abbett\inst{2}
\and {\AA}. Nordlund\inst{3}
}

\offprints{S.B.F. Dorch --- dorch@astro.su.se}

\institute{The Royal Swedish Academy of Sciences, 
           Stockholm Center for Physics, Astronomy and Biotechnology,
           SE-10691 Stockholm, Sweden
\and Space Sciences Laboratory, University of California, Berkeley, 
	CA-94720-7450, United States
\and The Niels Bohr Institute for Astronomy, Physics and Geophysics,
	Juliane Maries Vej 30, DK-2100 Copenhagen {\O}, Denmark
}

\date{Received date, accepted date}

\authorrunning{Dorch et al.}

\abstract{
We present 3-d numerical magneto-hydrodynamic simulations of a buoyant, twisted
magnetic flux rope embedded in a stratified, solar-like model convection zone.
The flux rope is given an initial twist such that it neither kinks nor
fragments during its ascent. Moreover, its magnetic energy content with respect
to convection is chosen so that the flux rope retains its basic geometry while 
being deflected from a purely vertical ascent by convective flows. The 
simulations show that magnetic flux is advected away from the core of the flux 
rope as it interacts with the convection. 
The results thus support the
idea that the amount of toroidal flux stored at or near the bottom of the 
solar convection zone may currently be underestimated.
}

\maketitle

\section{Introduction}

The concept of buoyant magnetic flux tubes is an essential part 
of the framework of current theories of dynamo action  
in stars, particularly in the case of cool dwarf stars such
as the Sun.
Results from studies of buoyant magnetic flux tubes 
carried out within the essentially 1-d thin flux tube approximation 
(e.g.\ 
Spruit \cite{Spruit1981} and 
Moreno-Insertis \cite{Moreno86})
are consistent with
the observed latitudes of emergence and tilt angles of
bipolar regions on the surface of the Sun 
(Fan et al. \cite{Fan+ea94} and Caligari et al. \cite{Caligari+ea95}).
More general 2-d simulations of flux tube cross-sections have shown that 
cylindrical tubes are disrupted by a magnetic Rayleigh-Taylor instability
(e.g.\ Sch\"{u}ssler \cite{Schussler1979}, Tsinganos \cite{Tsinganos1980}, 
Cattaneo et al. \cite{Cattaneo+ea90}, Matthews et al. \cite{Matthews+ea95},
Moreno-Insertis \& Emonet \cite{Moreno+ea96}). 
This renders them unlikely to reach the surface unless the presence of 
fieldline twist introduces a sufficient amount of magnetic tension to suppress 
this effect (Emonet \& Moreno-Insertis \cite{Emonet+ea96,Emonet+ea98} 
and Dorch \& Nordlund \cite{Dorch+ea98}). 
On the one hand, 3-d simulations of buoyant, twisted flux ropes have confirmed
several of the results from 2-d simulations
(Matsumoto et al. \cite{Matsumoto+ea98} and Dorch et al. \cite{Dorch+ea99}),
and have further shown that the S-shaped structure of a twisted flux tube as it 
emerges through the upper computational boundary is qualitatively similar to 
the sigmoidal structures observed in EUV and soft X-ray by the Yohkoh and SoHO 
satellites (e.g.\ 
Canfield et al. \cite{Canfield+ea99} and Sterling et al. \cite{Sterling+ea00}).
Moreover, tightly packed $\delta$-spots may be interpreted as the emergence of 
highly twisted, kinking flux ropes (e.g.\ Fan et al. \cite{Fan+ea99}).
On the other hand, it has been suggested that the value of the critical degree 
of twist needed to prevent the Rayleigh-Taylor instability may be 
unrealistically high in the 2-d case, and a smaller twist may be sufficient in 
the case of sinusoidal 3-d magnetic flux loops (Abbett et al. 
\cite{Abbett+ea00}).

In the solar convection
zone, buoyant flux structures are constantly interacting with the
surrounding convective downdrafts and updrafts, and 
the question remains whether the quasi-steady behavior that the 
flux ropes reach in the later phase of their rise in 2-d simulations 
(Emonet \& Moreno-Insertis \cite{Emonet+ea98} 
and Dorch \& Nordlund \cite{Dorch+ea98}) 
is stable towards perturbations from the surroundings, and whether the results 
found
for 3-d flux ropes moving in a essentially 1-d static
stratification are valid in the more realistic case. 

In this paper, we present our first results regarding the behavior of buoyant,
twisted flux ropes embedded in a fully dynamic model of solar-like convection.
 
\section{Numerical model}

The set-up of the model is twofold, consisting of a snapshot
of a time-dependent, but statistically relaxed ``local box'' convection zone 
model (sandwiched between two stable layers), and of an idealized twisted 
magnetic flux rope.  
We solved the full resistive and compressible MHD-equations
on a staggered 
mesh of 150 vertical $\times$ $105^2$ horizontal grid points, 
using the method by Galsgaard and others (e.g.\
Galsgaard \& Nordlund \cite{Galsgaard+ea97},
Nordlund et al.\
\cite{Nordlund+92}):
\begin{eqnarray}
 \frac{\partial \rho}{\partial t} &
             = & - \nabla \cdot \rho {\bf u}, \\
 \frac{\partial \rho {\bf u}}{\partial t} &
             = & - \nabla \cdot ( \rho {\bf u}{\bf u} - {\bf \tau})
                 - \nabla P + {\bf F}_{\rm grav} 
		+ {\bf F}_{\rm Lorentz}, \\
 \frac{\partial {\bf B}}{\partial t} &
             = & \nabla\times ({\bf u} \times {\bf B})
		+ \nabla \times (\eta \nabla \times {\bf B}), \\ 
 \frac{\partial e}{\partial t} &
             = & - \nabla \cdot (e {\bf u}) + P(\nabla \cdot {\bf u})
                 + Q_{\rm rad} + Q_{\rm visc} + Q_{\rm Joule}.
\end{eqnarray}
Here $\rho$, $\mathbf{u}$, $P$, $\mathbf{B}$, and $e$ represent the density, velocity, 
pressure, magnetic field, and internal energy respectively. $\eta$ and $\tau$ denote
the magnetic diffusivity and the viscous stress tensor; the source terms
$Q_{\rm visc}$, $Q_{\rm Joule}$, and $Q_{\rm rad}$ refer to the viscous, Joule, and
diffusive heating. In the upper part of the domain, $Q_{\rm rad}$ includes an additional
term that provides for a simple, isothermal cooling layer.

The computational method employs a finite difference staggered mesh with 
6th order derivative 
operators, 5th order centering operators, and a 3rd order 
time-stepping routine.
The diffusive terms are quenched in regions with smooth variations, 
to reduce the diffusion of well-resolved structures. 
Magnetic Reynolds numbers in non-smooth regions are of the order a few times $10^2$, 
but can be much higher in smooth regions.

\begin{figure}
\centering
\makebox[8cm]{
\vspace{0cm}
\hspace{0cm}\epsfxsize=4.5cm \epsfysize=4cm \epsfbox{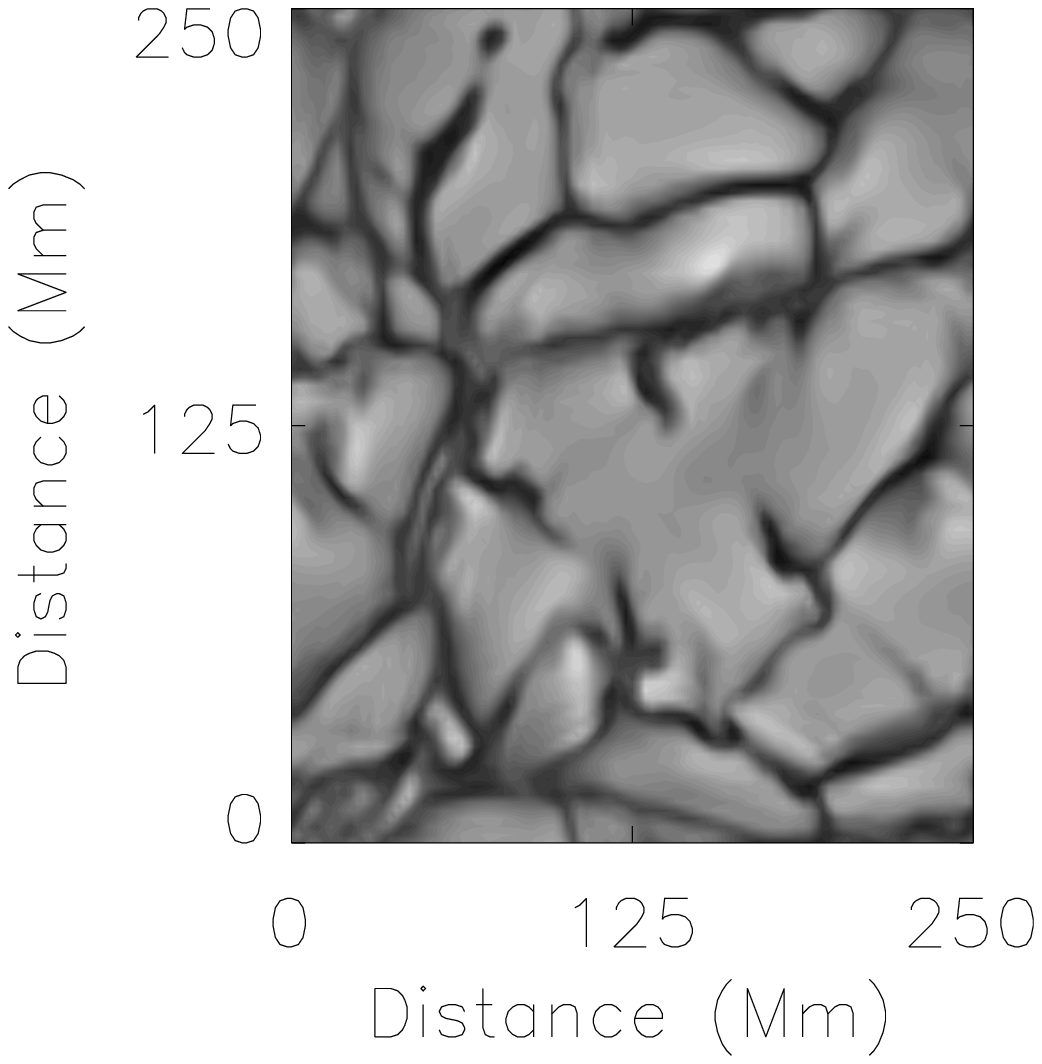}
\hspace{-0.75cm}\epsfxsize=4.5cm \epsfysize4cm \epsfbox{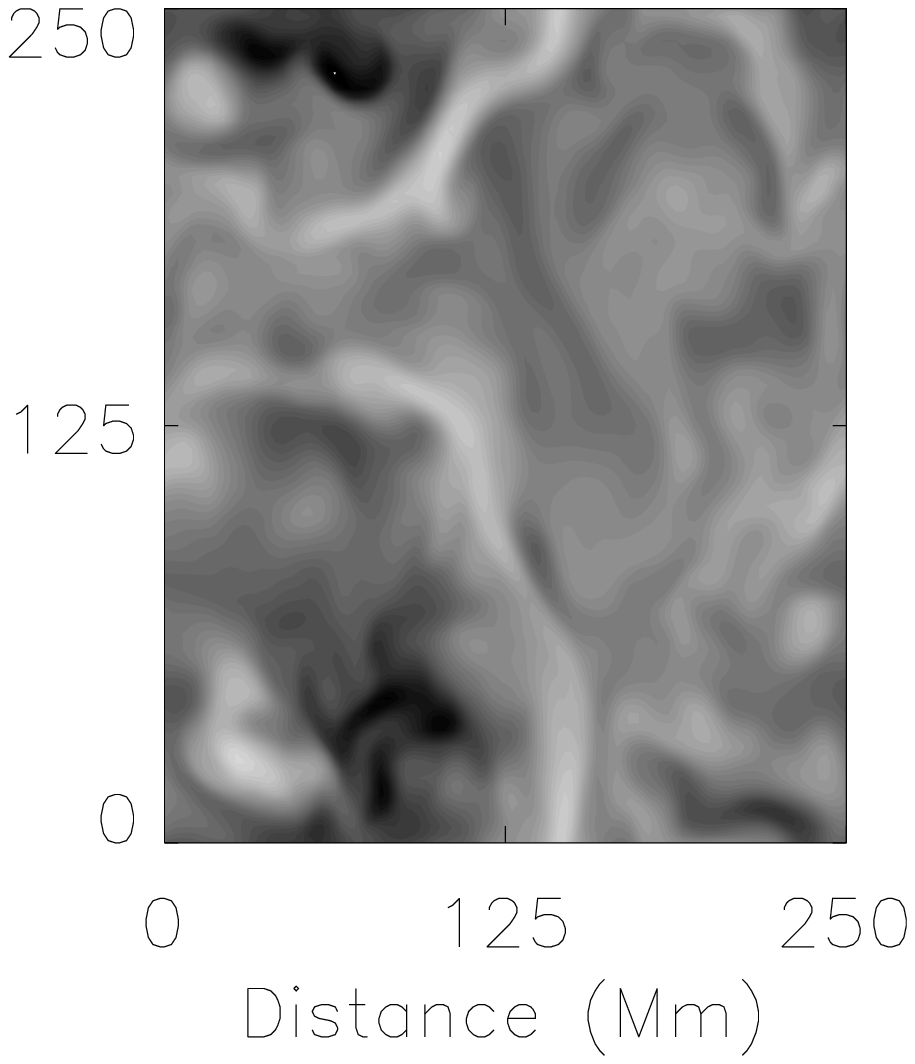}
\vspace{0cm} }
\caption[]{\small 
The vertical velocity in slices of the model convection 
zone in the initial state at two different depths: 
at the surface (left) and at the depth of the flux rope (right).
 } 
\label{fig1}
\end{figure}

The computational box is horizontally periodic with sides of 250 Mm and a
height of 313 Mm (of which 166 Mm is the convection zone, that covers 2.5
orders of magnitude in pressure).
The flows are turbulent through-out the convection zone, and the kinetic
energy spectrum displays a power law at intermediate wavenumbers
($k \approx$ 3 -- 10).
As it is typical for over-turning stratified convection, 
a cellular granulation pattern is generated on the surface of the 
convection zone (Fig.\ \ref{fig1}). The
typical length scale of this pattern is about 50 Mm, somewhat larger than
the canonical size of 32 Mm of solar super-granules
(e.g.\ Leighton et al. \cite{Leighton+ea62}).
The typical velocity of the granulation
is 200 m/s in the narrow downdrafts at the surface and slightly less 
in the upwelling regions, close to what is found for solar 
super-granulation (e.g.\ Worden \& Simon \cite{Worden+Simon76}).    

We choose an initially isentropic flux rope with 
a buoyancy of $1/\gamma$ (with $\gamma = 5/3$) lower than
the case of temperature balance, where the buoyancy is $1/\beta$
(with $\beta$ being the classical plasma beta). 
This is computationally advantageous, since we avoid the costly process of 
perturbing the flux rope from a state of mechanical equilibrium.

The initial twist of the flux rope is given by
\begin{equation}
 B_z = B_0 e^{-(r/{\rm R})^2}~ {\rm and}~ 
  B_{\phi} = \alpha (r/{\rm R}) B_z, \label{twist}
\end{equation}
where $B_z$ is the parallel and $B_{\phi}$ the transverse component 
of the magnetic field with respect to the rope's horizontal main axis. 
The coordinate system is chosen so that 
$z$ corresponds to the initial axial direction of the rope.

The wavelength $\lambda$ of the flux rope
is equal to the horizontal size of the domain 
so that $\lambda = 3.2~ {\rm H}_{\rm P0}$ 
at the initial position of the rope. 
Thus, the flux rope is not undular Parker-unstable even though the 
stratification permits this instability for longer wavelengths
(Spruit \& van Ballegooijen \cite{Spruit+ea82}).
The rope is initially twisted, with a pitch angle (at $r={\rm R}$) 
of $\psi_{\rm R}= \arctan (\alpha)$ and a 
radius ${\rm R}_0 = 0.177~ {\rm H}_{\rm P0}$
which corresponds to a half-width at half-maximum 
of $B_z$ (henceforth HWHM)
of $\sim 0.1~ {\rm H}_{\rm P0}$. 

To avoid problems associated with the large ratio of thermal to dynamic time 
scales, our convection model has a much higher luminosity than the Sun, and 
thus, all variables must be scaled to compare with solar values.
The choice of the magnetic field strength is somewhat problematic in this 
regard. 
The ratio of kinetic to thermal energy density $e_{\rm K}/e$ 
is much larger in our model than in the Sun (though the convective flows remain
subsonic with an average Mach number of 0.01). 
This requires a choice of $\beta$ that is smaller than its presumed value at 
the base of the solar convection zone so that the ratio of magnetic to kinetic 
energy density $e_{\rm M}/e_{\rm K}$ is the proper order of magnitude.
However, a small $\beta$ is what is needed so that the time it takes to 
complete a simulation is not prohibitively long.
We choose $\beta = 100$, which yields a solar-like $e_{\rm M}/e_{\rm K}$ of 100.

\section{Results}

We have performed several fully convective 3-d simulations, as well as 
a number of 2-d convection-less simulations. The results of the
latter agree with previous 2-d findings, and are used for reference
in the following. We discuss results from a 3-d simulation
with $\psi_{\rm R} = 45^{\circ}$ ($\alpha = 1$). 
In that case, the degree of twist is small enough to prevent the onset of the 
kink instability
(the linear growth rate vanishes for $\alpha = 1$, e.g.\ Fan et al. 
\cite{Fan+ea99}), 
yet it is large enough 
to prevent the onset of the Rayleigh-Taylor instability.
Thus, the rope retains its cohesion without distorting its shape
by any of these two instabilities, 
and we can focus our attention on the effects of the convective flows on 
the rope.  

\begin{figure*}[!htb]
\centering
\makebox[18cm]{
\vspace{0cm}
\hspace{0cm}\epsfxsize=5.6cm \epsfysize=5.6cm \epsfbox{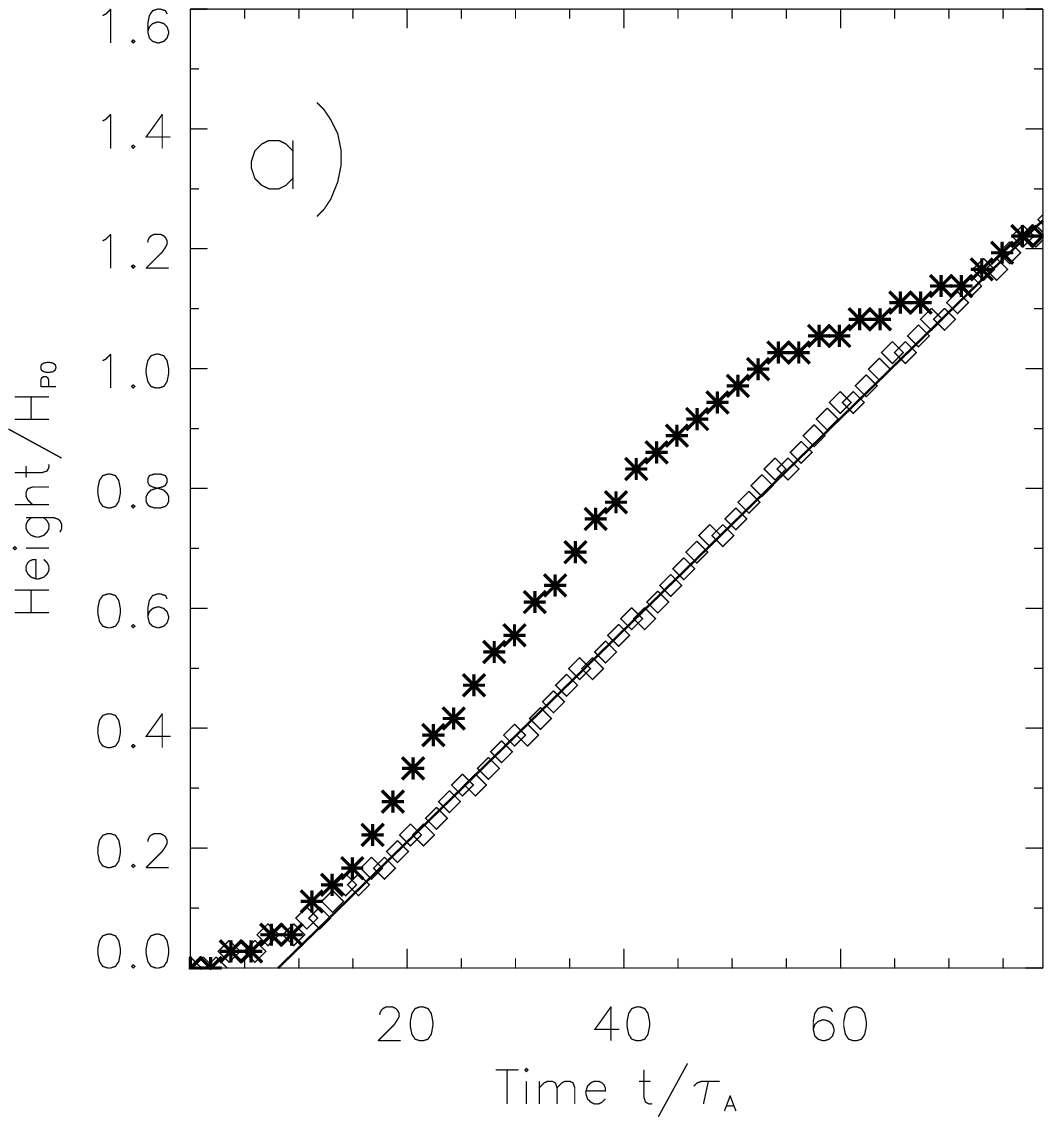}
\hspace{-1.55cm}\epsfxsize=2.8cm \epsfysize=5.6cm \epsfbox{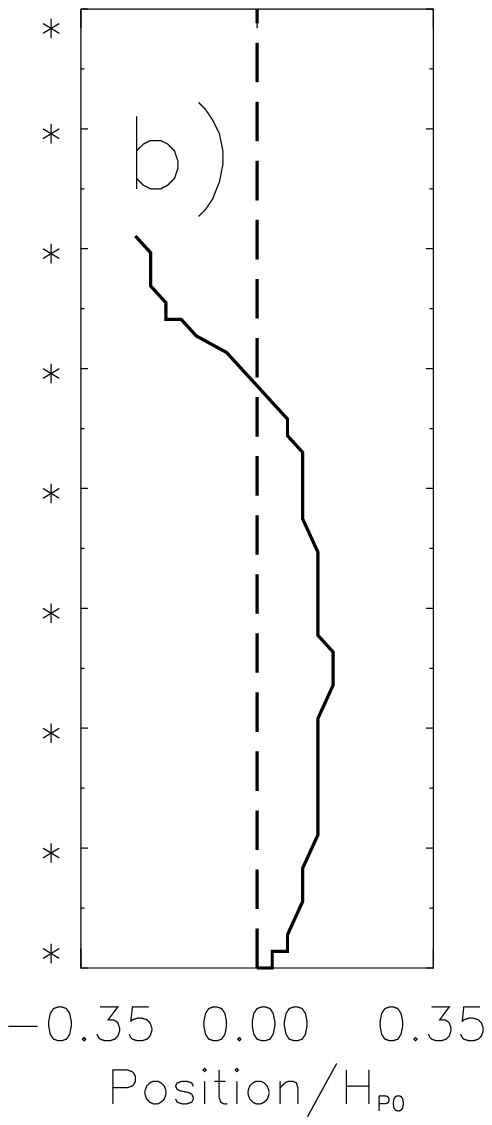}
\hspace{0cm}\epsfxsize=5.6cm \epsfysize=5.6cm \epsfbox{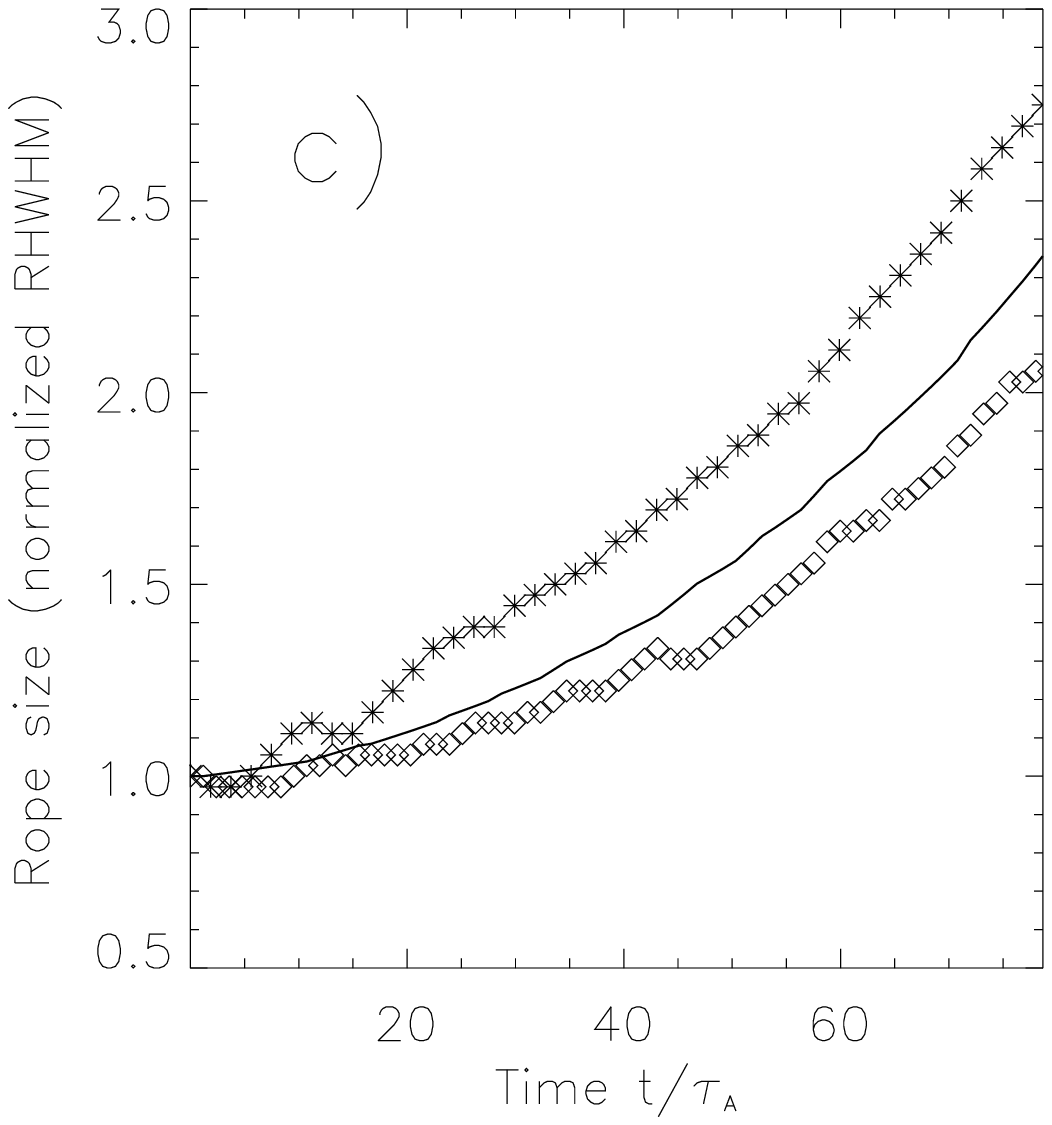}
\hspace{0cm}\epsfxsize=5.6cm \epsfysize=5.6cm \epsfbox{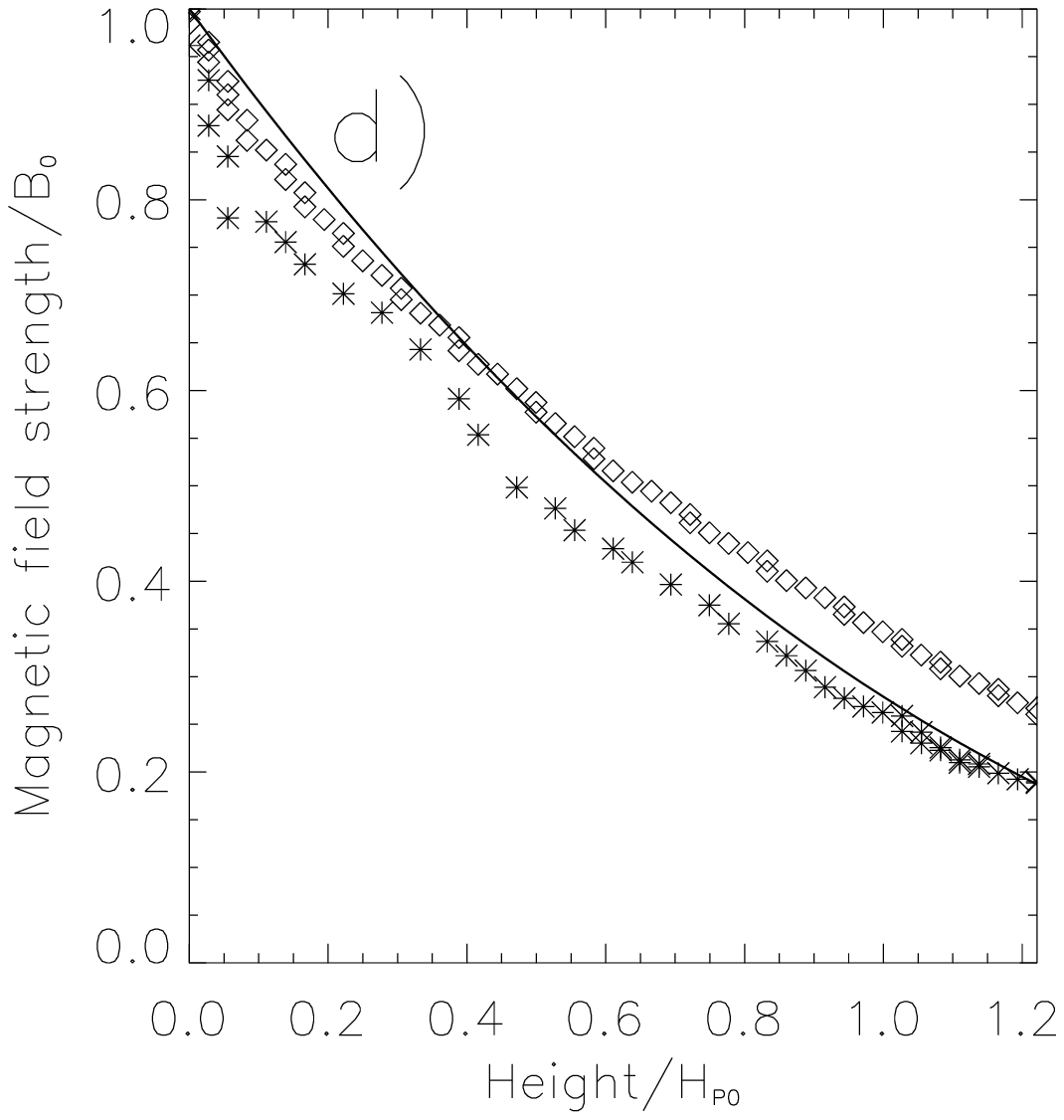}
\vspace{0cm} }
\caption[]{\small
 {\bf a}) height of the flux rope as a function of time (stars).
 Also plotted are the corresponding results from a 2-d reference simulation 
 (diamonds).
 The straight line corresponds to the average speed ($0.1~ v_{\rm A0}$) in 
 the rise phase.
 {\bf b}) drift of the flux rope in the meridional plane.
 {\bf c}) expansion of the flux rope (stars) with the corresponding result
  from the 2-d reference simulation over plotted (diamonds). 
  Also plotted is an analytical expression (solid line, see text). 
 {\bf d}) magnetic field strength as a function of height.}
 \label{fig2}
\end{figure*}

Fig.\ \ref{fig2} compares our 3-d simulation to a corresponding 2-d 
convection-less reference simulation, and a simple analytic flux tube.
As the 3-d rope rises, convective flows perturb its motion, preventing it 
from entering a well-defined terminal rise phase with a constant rise speed, 
as in the 2-d
reference simulation (see Fig.\ \ref{fig2}a). 
The rope remains straight and the maximum excursion of its axis, at the end of
the simulation, is $\sim 0.04~\lambda$, where we define the rope's axis as the
set of positions along the rope, where the magnetic field strength is maximum.
With the chosen super-equipartition axial field strength, the main action of 
the large-scale convective flows is to push the rope both left and right of 
the central plane (Fig.\ \ref{fig2}b; see also the mpeg-movie at Dorch 
\cite{Dorch-mpeg}), while the effect of the small-scale downdrafts 
(of the order of the rope's radius) is to locally deform its equipartition 
boundary.

Initially the rope is located in a general updraft region. This explains why 
the rise
speed of the rope is slightly greater than that of the 2-d reference simulation,
which reaches a terminal speed of $\sim 0.1~ v_{\rm A0}$ ($v_{\rm A0}$ being the
Alfv\'{e}n speed at the initial position of the rope). Nevertheless, the two
ropes arrive at the same final height at the end of the runs 
(though in the 3-d case, we note that the rope follows a longer path).
Our 3-d rope also expands more quickly than the rope in the 2-d simulation, 
and its rate of expansion is closer to what is expected 
from an adiabatically expanding, non-stretching tube with constant flux 
($B/\rho=$const.):
\begin{equation}
 {\rm R}(x) = {\rm R}_0 
   \left( 1 - \nabla_a \frac{x-x_0}{{\rm H}_{\rm P0}}\right)^{-1/2\nabla_a}. 
\label{analyt}
\end{equation} 
Fig.\ \ref{fig2}c shows the rope's characteristic size ${\rm R}_{hwhm}$
defined by the average between the vertical and horizontal HWHM along 
its axis 
(the short period oscillations due to differential buoyancy, that are not
well-resolved, have been filtered out by smoothing over two grid points).
As the rope rises and expands,
its magnetic field strength 
${\rm B}_c$, here defined as the average axial field strength along the rope, 
decreases at a rate close to that determined by Eq.\ (\ref{analyt}). 
At later times, the field strength of the 3-d and 2-d ropes
decrease at nearly the same rate (Fig.\ \ref{fig2}d).
The deviation can be attributed to the fact that, during its ascent, 
a significant
amount of the magnetic flux within the 3-d rope is lost to its surroundings.
\begin{figure}[!htb]
\centering
\makebox[9cm]{
\vspace{0cm}
\hspace{0cm}\epsfxsize=9.25cm \epsfysize=6cm \epsfbox{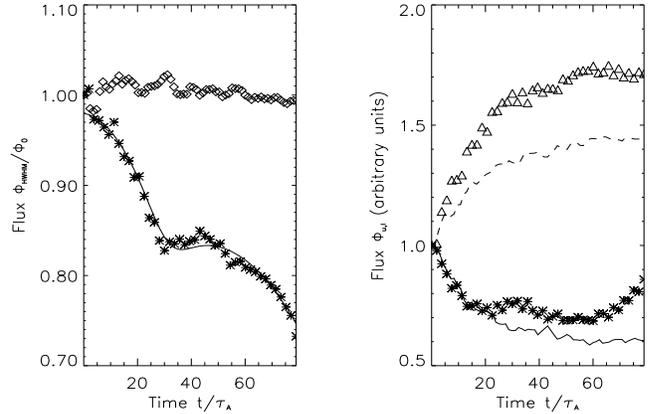}
\vspace{0cm} }
\caption[]{\small Left: magnetic flux within the
 rope $\Phi_i$ (stars), the corresponding quantity in the 2-d 
 simulation (diamonds) and an analytic fit (solid curve),
 Right: the normalized flux outside and above the center of the 3-d rope
 $\Phi_u$ (stars), and below, $\Phi_l$ (triangles). The same quantities are 
 shown for the 2-d reference 
 simulation (solid and dashed curves respectively). }
\label{fig3}
\end{figure}
This is illustrated by Fig.\ \ref{fig3} (left), which shows the total normalized magnetic
flux within the HWHM-boundary $\Phi_i$ as a function of time for both the 2-d
and 3-d ropes. We note that as the 3-d simulation progresses, the total flux-loss from 
the computational domain is only 0.3$\%$. 
The flux content of the rope, however, decreases much more quickly. 

Also shown in Fig.\ \ref{fig3} (right) is the magnetic flux external $\Phi_e$ 
to the rope both above and below its center $\Phi_u$ and $\Phi_l$ respectively. 
Since the sum $\Phi_e + \Phi_i$ is nearly conserved,
as $\Phi_i$ decreases, $\Phi_e = \Phi_u + \Phi_l$ must increase by an equal 
amount. 
However, the distribution of the flux-loss is not symmetric: more flux is lost 
to the surroundings below the rope than above. 
The e-folding time of the increase of
flux $\Phi_l$ in the lower domain is $\sim 20~ \tau_{\rm A}$ 
(with $\tau_{\rm A} = {\rm R}_0/v_{\rm A0}$).

This asymmetry also exists in the 2-d reference simulation, even though the 
total flux-loss is much smaller in that case. 
The asymmetry is a result of two factors. First, as the rope rises, the total 
volume 
above it decreases, while the volume below it increases. Second, there is an 
anti-symmetry of the relative velocity across the rope.
When the rope ascends, there is a tendency for flux to be advected towards the 
rope near its apex, and transported away from the rope in its wake.
The more pronounced asymmetry in the 3-d case
can be attributed to the pumping effect that transports the weak
field downwards (Dorch \& Nordlund \cite{Dorch+Nordlund01} and
Tobias et al.\ \cite{Tobias+ea01}). 

We have defined the flux rope as the magnetic structure
that lies within the HWHM-boundary. This boundary is not, however, a contour 
that moves with the fluid in the classical sense: the flux within the latter
kind of contour is naturally conserved (neglecting resistive effects) and
equal to the total flux $\Phi_0 = \Phi_e + \Phi_i$. 
The HWHM-boundary is a convenient way of defining the flux rope and a
characteristic size 
${\rm R}_{hwhm}$,
that behaves more or less
as it is expected from the analytical expression Eq.\ (\ref{analyt}).
The evolution of the flux within the rope's core $\Phi_i$ is determined by
\begin{equation} 
 \dot{\Phi}_i = - \oint_{\rm boundary} \Delta {\bf v}\times 
  {\bf B} \cdot {\rm d}{\bf l}, \label{dotflux1}
\end{equation} 
where $\Delta {\bf v}$ is the difference between the fluid
velocity and the motion of the HWHM-boundary. 
The average ``slip'' $\overline{\Delta v}$ of the rope's boundary in the simulation
is only a small fraction of the rise speed, and varies between the range of 
plus or minus a few times $10^{-4}~$ and $10^{-3}~ v_{\rm A0}$.

Making the rather crude
assumptions that the boundary only moves radially relative to the fluid
and that the circumference of the boundary is circular (which it is not),
Eq.\ (\ref{dotflux1}) reduces to
\begin{equation} 
 \dot{\Phi}_i = - \pi {\rm R}_{hwhm}~ \Delta v~ {\rm B}_c, \label{dotflux2}
\end{equation} 
where ${\rm B}_c$ is the field strength at the center of the flux rope.
Integrating Eq.\ (\ref{dotflux2}) numerically with 
${\rm R}_{hwhm}$ and ${\rm B}_c$
determined from the simulation (see Fig.\ \ref{fig2}), the result is a perhaps 
surprisingly good fit to the actual flux-loss, see Fig.\ \ref{fig3} (left),
if $\Delta v$ is set to 3~10$^{-4}~ v_{\rm A0}$
throughout the time span of the simulation except for a short interval of
$\sim 3~ \tau_{\rm A}$ around $t = 30~ \tau_{\rm A}$,
where $\Delta v = - 10^{-3}~ v_{\rm A0}$,
when the rope passes from one updraft to another (see the discussion below).

\section{Discussion and conclusions}

The 3-d numerical simulations show that the interaction of a buoyant twisted
flux rope with stratified convection leads to a considerable loss of magnetic
flux from the core of the flux rope (as defined by the rope's HWHM-boundary).  

The initial position of a flux rope in the convection zone is 
significant for the 
subsequent detailed history of its rise: with the present convective 
flows and the initial location of the flux rope, 
most of the rope starts out located inside or close to a convective updraft. 
Thus, the ascent of the rope is likely to be influenced by this fact, and
we are therefore not able to draw any conclusions on the detailed path of its rise.
However, in the course of the simulation, the flux rope rises 96 Mm, and 
loses about 25\% of its original flux content. 
This, ceteris paribus, leads to an increase in the 
amount of toroidal flux that must be stored at the bottom 
of the convection zone during the course of the solar cycle. 

In the Sun, toroidal flux ropes rise about 200 Mm through the convection zone 
before
emerging as bipolar active regions. One may thus expect them to lose even
more of their initial flux, which would then be pumped back down toward the
bottom of the convection zone. We can quantify
this subsequent flux-loss by assuming that Eq.\ (\ref{dotflux2}) is valid 
through-out the rise, that the ropes expand according to the simple
analytical estimate of Eq.\ (\ref{analyt}), and that the ratio
of the slip $\Delta v$ to the rise speed remains constant. 
Given these assumptions,
the flux-loss at a height of 200 Mm is 26 \% of the initial flux, i.e.\
not much more than in our simulation. 
However, 
the relative slip may not remain constant throughout the rope's rise. 
For example, $\Delta v$ and thus $\dot{\Phi}_i$ changes at the time
around $t=30~ \tau_{\rm A}$, which corresponds to the time when the rope 
is at its
maximum (rightward) excursion from a vertical ascent (see Fig.\ \ref{fig2}b).
At that time, the rope exits the convective updraft with which it
was initially associated, and enters a different ascending ``plume'' to
the left of the its original position. 
This leads to a transient compression of the
rope ($\Delta v < 0$, in the simplified expression Eq.\ \ref{dotflux2}).
After entering the new plume, the average slip returns to its previous 
positive value for the remainder of the rise. 

Petrovay \& Moreno-Insertis (\cite{Petrovay+Moreno97}) suggested that 
turbulent erosion of magnetic flux tubes may take place within the solar
convection zone due to the ``gnawing'' of turbulent convection. They
propose a mechanism whereby a flux tube is eroded by a thin current
sheet that forms spontaneously within a diffusion time.
That we do not see a loss of flux via this type
of enhanced diffusion should not, however, be taken as a dis-proof of the 
feasibility of turbulent erosion: it requires the turbulence to be 
resolved down to much smaller scales $\ell \ll \lambda$, than in our
simulations.
Instead, the flux-loss
is completely due to the advection of flux away from the core of the flux rope
by convective motions. Most of the flux that is ``gnawed-off'' ends up in
the trailing wake and some of this flux is mixed back into the upper layers
by ascending flows. We speculate that both types of flux-loss may take place 
simultaneously in the Sun, and as a result, the amount of toroidal 
flux stored near the bottom of the solar convection zone may 
currently be underestimated.

\begin{acknowledgements}
SBFD and BVG was supported through an EC-TMR grant to the European
Solar Magnetometry Network. 
WPA was supported through the NSF and NASA's SECT and Solar Physics Research
and Analysis programs.
Computing time was provided by the Swedish National Allocations Committee.
The authors thank Krist\'{o}f Petrovay for discussions on 
flux-loss mechanisms.
 
\end{acknowledgements}

\end{document}